\documentclass[10pt,letter,twoside]{IEEEtran}
\pdfoutput=1

\usepackage{ifpdf}              % Allows for \ifpdf conditions
\usepackage{hyperref}           % Nicely hyperlinked text (needs pkg ifpdf)
\usepackage{cite}               % Gets the IEEE citation numbers right
\usepackage[cmex10]{amsmath}    % Powerfull commands for dealing with math
\usepackage{array}              % Improves LaTeX2e array and tabular environments
\usepackage{amsfonts}
\usepackage{amssymb}
\usepackage{xcolor}

\usepackage{dsfont}
\usepackage{mathtools}
\usepackage{algorithm}
\usepackage{algpseudocode}
\usepackage{amsthm}

\ifpdf
\usepackage[pdftex]{graphicx}
\usepackage[font=small]{caption}
\usepackage{subcaption}
\usepackage{balance}

\usepackage{epstopdf}
 \DeclareGraphicsExtensions{.eps,.jpg,.png,.pdf}
\else
 \usepackage[dvips]{graphicx}
 \DeclareGraphicsExtensions{.eps}
\fi

\makeatletter
\def\BState{\State\hskip-\ALG@thistlm}
\makeatother

\usepackage{lipsum}
\renewcommand{\P}[1]{P_\text{#1}}

\newcommand{\E}[1]{\mathbb{E} \left\{#1 \right\}}

\newcommand{\I}[1]{\mathds{1}\left\{ #1 \right\}}
\renewcommand{\P}[1]{\mathbb{P}\left\{ #1 \right\} }
\newcommand{\Pw}[2]{\mathbb{P}_{#1}\left\{ #2 \right\} }

\newcommand{\fig}[3]
{
  \begin{figure}[!t]	
    \centering
    \includegraphics[width=3.5in]{#1}
    \caption{#3}
    \label{#2}
  \end{figure}
}

\ifCLASSOPTIONpeerreview

\else

\fi

\title{Social Learning Against Data Falsification in Sensor Networks}

\ifCLASSOPTIONjournal

\author{
	Fernando Rosas and Kwang-Cheng Chen, \IEEEmembership{Fellow, IEEE}
        \thanks{The authors are with the Graduate Institute of Communication Engineering, National Taiwan University, Taipei 10617, Taiwan (email \{nrosas, ckc\}@ntu.edu.tw). K.C. Chen is also with the Department of Electrical Engineering, University of South Florida, Tampa, FL 33620, USA (email kwangcheng@usf.edu).}
 }

\fi

  \ifCLASSOPTIONpeerreview  % Peer-review mode
\author{
	Fernando Rosas$^1$ and Kwang-Cheng Chen$^{1,2}$, \IEEEmembership{Fellow, IEEE} \\
	$^1$ Graduate Institute of Communication Engineering, National Taiwan University\\
        $^2$ Department of Electrical Engineering, University of South Florida\\
        Email: \{nrosas, ckc\}@ntu.edu.tw, kwangcheng@usf.edu
}

\fi
\begin{document}

% ---------------------------------------------------------------- %
% --- TITLE ------------------------------------------------------ %
% ---------------------------------------------------------------- %
\maketitle

% ---------------------------------------------------------------- %
% --- ABSTRACT --------------------------------------------------- %
% ---------------------------------------------------------------- %
\begin{abstract}
Although surveillance and sensor networks play a key role in Internet of Things, sensor nodes are usually vulnerable to tampering due to their widespread locations. In this letter we consider data falsification attacks where an smart attacker takes control of critical nodes within the network, including nodes serving as fusion centers. In order to face this critical security thread, we propose a data aggregation scheme based on social learning, resembling the way in which agents make decisions in social networks. Our results suggest that social learning enables network resilience, even when a significant portion of the nodes have been compromised by the attacker. Finally, we show the suitability of our scheme to sensor networks by developing a low-complexity algorithm to facilitate the social learning data fusion rule in devices with restricted computational power. 
\end{abstract}


% Put extra information on the cover page in peer-review mode and insert a page break and the second title (ignored for other modes).
%\ifCLASSOPTIONpeerreview
%\begin{center}
%Corresponding author: Fernando Rosas, \texttt{ferosas@puc.cl}.\\
%\end{center}
%\fi
%\IEEEpeerreviewmaketitle

% ---------------------------------------------------------------- %
% --- MAIN TEXT -------------------------------------------------- %
% ---------------------------------------------------------------- %

\section{Introduction}
\label{intro}

Large distributed sensor networks typically provide surveillance services over extensive areas, such as activity monitoring in military or secure zones, protection of drinkable water tanks from chemical attacks, or intrusion detection~\cite{veeravalli2012distributed,barbarossa2013distributed}. However, the reliability of these networks is in many cases limited due to the high vulnerability of the sensor nodes~\cite{shi2004designing}. In reality, nodes are frequently located in unprotected locations and are susceptible to physical or cyber captures. Moreover, nodes are generally not tamper-proof due to cost concerns, and their limited computing power, memory, and energy capabilities do not allow sophisticated cryptographic techniques.

One serious threat to the reliability of distributed surveillance is the data falsification or ``Byzantine'' attack, where an adversary takes control over a number of authenticated nodes~\cite{marano2009distributed}. Following the classic \textit{Byzantine Generals Problem} \cite{lamport1982byzantine}, Byzantine nodes can generate false sensing data, exhibit arbitrary behaviour or collude in order to create a networked malfunction. The effect of data falsification attacks over distributed detection has been intensely studied, characterizing the impact over the detection performance and also proposing various defense mechanisms (c.f. \cite{vempaty2013distributed} for an overview, and also \cite{6778791,6998053,7134807}). These works focus in networks with star or tree topology, where the data is gathered in a special node called ``fusion center'' (FC) that is responsable for the final decision.%. kailkhura2014distributed,,,,7065266

%awerbuch2004mitigating

A key assumption in the literature is that the adversary can compromise regular sensor nodes but not the FC itself. However, in many scenarios the short range of the nodes' transmissions force the FC to be installed in unsafe locations, being vulnerable to tampering as well. A tampered FC completely disables the detecting capabilities of the network, generating a single point of failure and hence becoming the weakest point of the system~\cite{parno2005distributed}. To address this serious security thread, this letter is novel in considering powerful topology-aware data falsification attacks, where the adversary knows the network topology and leverage this knowledge to take control of the most critical nodes of the network ---either regular nodes or FCs.  This represents a worst-case scenario, where the network structure has been disclosed e.g. from network tomography via traffic analysis\cite{castro2004network}.

The design of reliable distributed detection schemes is a challenging task. In effect, even though the distributed sensing literature is vast (see e.g. \cite{veeravalli2012distributed,barbarossa2013distributed} and references therein), the construction of optimal schemes is in general NP-hard \cite{tsitsiklis1985complexity}. Moreover, although in many cases the optimal schemes can be characterized as a set of thresholds for likelihood functions, the determination of these thresholds is usually an intractable problem~\cite{tsitsiklis1993decentralized}. For example, symmetric thresholds can be suboptimal even for networks with similar sensors arranged in star topology~\cite{warren1999optimum}, being only asymptotically optimal when the network size increases \cite{tsitsiklis1993decentralized,chamberland2004asymptotic}. Moreover, symmetric strategies are not suitable for more elaborate network topologies, and hence heuristic methods are usually necessary.

To deal with this dilemma, in this letter we propose a low-complexity data aggregation scheme based on \textit{social learning} principles, which resembles social decisions-making processes while avoiding fusion center functions~\cite{bikhchandani1992theory,acemoglu2011bayesian,krishnamurthy2013social}. The scheme is a threshold-based data fusion strategy related to the ones considered in~\cite{tsitsiklis1993decentralized}. However, its connection with social decision-making enables an intuitive understanding of its inner mechanisms, and also allows an efficient implementation that is suitable for the limited computational capabilities of a sensor node. For avoiding the security threads introduced by fusion centers, our scheme uses a tandem or serial topology~\cite{viswanathan1988optimal,papastavrou1992distributed,swaszek1993performance,viswanathan1997distributed,bahceci2005serial}. Contrasting with the literature, our analysis does not focus on optimality issues of the data fusion but aims to illustrate how this scheme can enable network resilience against a powerful topology-aware data falsification attacker, even when a significant number of nodes have been compromised. %easley2010networks

\section{System model and problem statement}
\label{sec:2}

\subsection{System model}

We consider a network of $N$ sensor nodes that are deployed over an area where surveillance is needed. The output of the sensor of the $n$-th node is denoted by $S_n$, taking values over a set $\mathcal{S}$ that can be discrete or continuous. Based on these signals, the network needs to infer the value of the binary variable $W$, with events $\{W=1\}$ and $\{W=0\}$ corresponding to the presence or absence of an attack, respectively. No knowledge about of the prior distribution of $W$ is assumed, as attacks are rare and might follow unpredictable patters.

We consider nodes with equal sensing capabilities, and hence assume that the signals $S_n$ are identically distributed. For the sake of tractability, it is assumed that the variables $S_1,\dots, S_N$ are conditionally independent\footnote{The conditional independency of sensor signals is satisfied when the sensor noise is due to local causes (e.g. thermal noise), but do not hold when there exist common noise sources (e.g. in the case of distributed acoustic sensors~\cite{bertrand2011applications}).} given the event $\{W=w\}$, following a probability distribution denoted by $\mu_w$. It is assumed that both $\mu_0$ and $\mu_1$ are absolutely continuous with respect to each other~\cite{Loeve1978}, i.e. no particular signal determines $W$ unequivocally. The log-likelihood ratio of these two distributions is therefore given by the logarithm of the corresponding Radon-Nikodym derivative $\Lambda_S(s) = \log \frac{d \mu_1}{d \mu_0} (s) $\footnote{When $S_n$ takes a finite number of values then $\frac{d \mu_1}{d \mu_0} (s) = \frac{ \P{ S_n=s|W=1}}{ \P{ S_n=s|W=0}}$, while if $S_n$ is a continuous random variable with conditional p.d.f. $p(S_n|w)$ then $\frac{d \mu_1}{d \mu_0} (s) = \frac{ p(s|w=1) }{ p(s|w=0) }$.}. %It is also assumed that $\mu_0 \neq \mu_1$, so that $\Lambda(S_n)$ is not trivially equal to zero.

In addition to sensing hardware, each node is equipped with computing capability and a low-power transceiver to transit and receive data. However, battery limitations impose severe restrictions over the communication bandwidth, and thus it is assumed that each node forward its data to others by broadcasting a binary variable $X_n$. Note that these signals could be appended to wireless control packages and viceversa.%, or also could be shared by light, ultrasound or other media.

%Without the loose of generality, t
The nodes transmit their signals sequentially according to their indices. 
% (i.e. node 1 transmits first, then node 2, etc).   
Due to the nature of wireless broadcasting, which might be overlooked in some security literatures, nearby transmissions can be overheard. 
%This letter focuses on the case where each node can access  
%This letter focuses on the case of a fully connected network%\footnote{This assumption is satisfied in environment with benign propagation properties, like open landscapes (e.g. desertic areas).}
Therefore, it is assumed that 
the $n$-th node can generate $X_n$ based on information provided by $S_n$ and $\boldsymbol{X}^{n-1}=(X_1,\dots,X_{n-1})$. 
%Note that, with this assumption, we are leaving routing and other issues that arise in multihop networks for future work.  
A \textit{strategy} %$\{\pi_n\}_{n=1}^N$ 
is a collection of functions $\pi_n:\mathcal{S}\times \{0,1\}^{n-1}\to \{0,1\}$ such that $X_n = \pi(S_n,\boldsymbol{X}^{n-1})$. 
Although the burden of overhearing all the previously broadcasted signals can be reduced by designing smart network topologies and routing strategies, these networking functions are left for future studies.

The network operator collects the transmitted packages from a specific node labeled as $n_\text{c} \in \{1,\dots,N\}$, possibly employing unmanned ground or aerial vehicles that access a shared signal at a specific network location, or by using a shared communication channel. 
%Therefore, $X_{n_\text{c}}$ constitutes the output of the distributed detection scheme. 
The network performance is quantified by the corresponding miss-detection and false alarm rates, given by $\P{\text{MD}} = \P{ X_{n_\text{c}} = 0 | W=1}$ and $\P{\text{FA}}  = \P{ X_{n_\text{c}} = 1 | W=0}$, respectively.

Finally, it is assumed that $N^*$ Byzantine nodes are controlled by an adversary without being noticed by the network operator. The adversary can freely define the values of the binary signals transmitted by byzantine nodes in order to degrade the network performance. It is further assumed that the adversary is ``topology-aware'', knowing the node sequence and the strategy %$\{ \pi_n\}_{n=1}^N$ 
that is in use. Therefore, the adversary could well control the $N^*$ most critical nodes in terms of network performance. However, the adversary has no knowledge about $n_\text{c}$, as it can be chosen at run-time and changed regularly. 

\vspace{-0.2cm}

\subsection{Problem statement}
\label{qweqw213123}

Our goal is to develop a network-resilient strategy to mitigate the effect from a powerful topology-aware adversary when the network operator (i.e. defender) has no knowledge of the number of Byzantine nodes or other attack's statistics. Note that in most surveillance applications miss-detections are more important than false alarms, being difficult to estimate the cost of the worst-case scenario. Therefore, the system performance is evaluated following the Neyman-Pearson criteria by setting an allowable false alarm rate and focusing on the achievable miss-detection rate. %Note than finding optimal solutions is a formidable problem, even for the simple case of networks with start topology and no byzantine attacks~(see \cite{chamberland2007wireless} and references therein). 
%~\cite{smith2011network}

%Please note that m
Most signal processing techniques for distributed detection rely on a FC(s) that gather data and generate estimators, and sensor nodes that provide informative signals to them~\cite{rajagopalan2006}. 
Intuitively, if $X_n$ is influenced by $X_m$ with $m<n$, this would ``double-count'' the information provided by $S_m$. Therefore, in order to guarantee diversity, traditional distributed detection schemes choose to ignore previously broadcasted signals.
%This leads to good performance statistics, achieving exponentially decaying miss-detection rates with respect to the number of sensing nodes~\cite{chamberland2003decentralized}.
However, as nodes don't perform any data aggregation, each of their shared signals are not, by themselves, good estimations of the target variable. This generates a single point of failure in the network, as if the adversary compromises the FC(s) then the only accurate estimator that exist within the network is lost and hence the inference process fails. %Note that the low-power transceivers of wireless sensing nodes usually don't transmission ranges beyond 40 meters, and therefore it is likely that the fusion center may also be in a vulnerable location and hence be victim of tampering. %The FC is the most critical node to the detection performance, and therefore would be the most endangered element of the network. 

\section{Social learning as a data aggregation scheme}
\label{sec:III}

%This section introduces a data aggregation scheme inspired by social learning principles, and explains its functions against topology-aware data falsification attacks. In the sequel, Section~\ref{secA} analyses the data fusion rule; then Section~\ref{qweqwe234153} presents a low-complexity implementation; and finally Section~\ref{secC} discusses critical aspects of the proposed scheme.

\subsection{Data fusion rule}
\label{secA}

Social learning models supply new directions to analyze the sequential decision processes where agents combine personal information and peers' opinions~\cite{krishnamurthy2013social}. Applied to a sensor network, each node can be considered as an agent that decides the presence of attacks based on measurements and overheard signals from other nodes. In this letter we consider rational agents that follow a \textit{Bayesian strategy}, denoted as $\pi_n^\text{b}(S_n,\boldsymbol{X}^{n-1})$, which can be described by
\begin{equation}\label{eq:bayes}
\frac{ \P{W = 1|S_n,\boldsymbol{X}^{n-1}} }{ \P{W = 0|S_n,\boldsymbol{X}^{n-1}} }
\mathop{\lessgtr}_{\pi_n^\text{b}=1}^{\pi_n^\text{b}=0}
\frac{ u(0,0) - u(1,0)  }{ u(1,1) - u(0,1) }
\enspace.
\end{equation}
Above, $u(x,w)$ is a cost assigned to the decision $X_n=x$ when $W=w$, which can be engineered in order to match the relevance of miss-detections and false alarms~\cite{poor2013introduction}. Moreover, by noting that $X^{n-1} = \pi_{n-1}^\text{b}(S_{n-1},\boldsymbol{X}^{n-2})$ is influenced only by $S_1,\dots,S_{n-1}$, the conditional independency of the signals imply that $S_n$ and $\boldsymbol{X}^{n-1}$ are also conditionally independent given $W=w$. Therefore, using the Bayes rule, a direct calculation shows that \eqref{eq:bayes} can be re-written as
\begin{equation}\label{eq:asdasdA3}
\Lambda_S(S_n) + \Lambda_{\boldsymbol{X}^{n-1}}(\boldsymbol{X}^{n-1}) \mathop{\lessgtr}_{\pi_n^\text{b}=1}^{\pi_n^\text{b}=0} \tau
\enspace,
\end{equation}
where $\tau =  \log \frac{ \P{W=0} }{ \P{W=1} } + \log \frac{ u(0,0) - u(1,0) }{ u(1,1) - u(0,1)  }$ and $\Lambda_{\boldsymbol{X}^{n-1}}(\boldsymbol{X}^{n-1}) $ is the log-likelihood ratio of $\boldsymbol{X}^{n-1}$. As the prior distribution of $W$ is usually unknown, the network operator needs to select the lowest value of $\tau$ that satisfies the required false alarm rate given by the Neyman-Pearson criteria (c.f. Section~\ref{qweqw213123}).

As in a realistic scenario the statistical properties of the potential topology-aware data falsification attacks are not available to the defender, our approach is to make each node to follow a bayesian strategy ignoring the potential attack. Such an approach has three attractive features: 
\begin{itemize}
\item[1.] Provides a computation rule that does not need to adapt to different attacker's profiles.
\item[2.] Minimizes the average cost $\E{ u(\pi_n(S_n,\boldsymbol{X}^{n-1}),W) }$ when no attacks take place \cite{poor2013introduction}.
\item[3.] Enables network resilience (c.f. Section III-C and IV).
\end{itemize}

%This condition provides a simple data fusion rule to combine the information provided by $S_n$ and $\boldsymbol{X}^{n-1}$.

%In effect, a high value of $\tau$ in \eqref{eq:asdasdA3} makes easier to decide in favour of $\{X_n=0\}$, reducing the false alarms at the cost of more miss-detections, while a low value of $\tau$ has the opposite effect. Note that a minimax solution~\cite{poor2013introduction} might not guarantee optimality in this scenario, as it is the Neyman-Pearson criterion ---and not the average cost--- what leads the system optimization.

%It is important to notice that, although the above derivation is based on the minimization of the average error rate, our interest is to use \eqref{eq:asdasdA3} in a scenario that follows the Neyman-Pearson criteria. Our choise of data fusion rule is motivated by it simplicity, and due to the fact that threshold rules over likelihood functions are also optimal under the Neyman-Person formulation \cite{tsitsiklis1993decentralized}.

Clearly Byzantine nodes do not follow \eqref{eq:asdasdA3}, as their interest is to degrade the network performance. Let us denote as $\mathcal{B}$ the set of indices of the Byzantine nodes and $N^*$ the cardinality of $\mathcal{B}$. As events $\{W=0\}$ are much more frequent than $\{W=1\}$, any abnormal increase of the false alarm rate would be easily noted and hence provides no benefit to the adversary. Therefore, a rational strategy for the adversary is to increase the miss-detection rate by forcing $X_n=0$ for all $n \in\mathcal{B}$.

\subsection{An algorithm for computing the social log-likelihood}
\label{qweqwe234153}
The only challenge for implementing \eqref{eq:asdasdA3} in a sensor node as a data fusion rule is to have an efficient algorithm for computing  $\Lambda_{\boldsymbol{X}^{n-1}}(\boldsymbol{x}^{n-1}) $. For finding such an algorithm, a direct application of the chain rule of probabilities shows that
\begin{equation}
\Lambda_{\boldsymbol{X}^n}(\boldsymbol{x}^{n}) = \log \prod_{k=1}^{n} \frac{ \P{ X_k =x_k | \boldsymbol{X}^{k-1} = \boldsymbol{x}^{k-1},W=1 } }{ \P{X_k=x_k | \boldsymbol{X}^{k-1} =\boldsymbol{x}^{k-1},W=0 }}, \nonumber
\end{equation}
with the understanding that $\boldsymbol{X}^{0} = \boldsymbol{x}^{0}$ is null. Then, following the discussion presented in Section~\ref{secA}, we compute $\mathbb{P}\{ X_k = x_k \,|\, \boldsymbol{X}^{k-1} = \boldsymbol{x}^{k-1},W=w \}$ ignoring potential attacks. Assuming that the $k$-th node is not a Byzantine node, one obtains
\begin{align}
\mathbb{P}\{ &X_k = 0 \,|\, \boldsymbol{X}^{k-1} = \boldsymbol{x}^{k-1}, W=w \}  \nonumber \\
&= \int_\mathcal{S} \P{ X_k = 0 | {\boldsymbol{X}^{k-1}}=\boldsymbol{x}^{k-1}, W=w,S_k=s } \text{d} \mu_w(s) \nonumber\\
&= \int_\mathcal{S} \I{ \pi_k^\text{b}(s,{\boldsymbol{x}^{k-1}}) = 0 } \text{d} \mu_w(s) \nonumber\\
&= \Pw{w}{ \Lambda_S(S_k) + \Lambda_{\boldsymbol{X}^{k-1}}({\boldsymbol{x}^{k-1}}) < \tau} \nonumber\\
&= F_w^\Lambda(\tau - \Lambda_{\boldsymbol{X}^{k-1}}(\boldsymbol{x}^{k-1})) \label{eqL21344315}
\enspace,
\end{align}
where $F_w^\Lambda(\cdot)$ is the c.d.f. of the variable $\Lambda_s(S_n)$ conditioned to $W=w$. 
%Note that the first equality is a consequence of the conditional independency between $S_k$ and $\boldsymbol{X}^{k-1}$ given $W=w$. 
Using the above results, it can be shown that
\begin{equation}
\Lambda_{\boldsymbol{X}^{n+1}}(\boldsymbol{x}^{n+1}) - \Lambda_{\boldsymbol{X}^n}(\boldsymbol{x}^{n}) = \lambda(x_k, \tau - \Lambda_{\boldsymbol{X}^{n}}( \boldsymbol{x}^{n})) \nonumber %\\
\enspace,
\end{equation}
where $\lambda(\cdot,\cdot)$ is defined as
\begin{equation}\label{eq:firstll}
\lambda(x,a) = x \log \frac{ F_1^\Lambda(a) } { F_0^\Lambda (a) }
+
(1-x) \log \frac{ 1- F_1^\Lambda (a)} { 1 - F_0^\Lambda (a) } \nonumber
\enspace.
\end{equation}

Leveraging above derivations, we develop Algorithm 1 as a simple iterative procedure for computing $\Lambda_{\boldsymbol{X}^n}(\boldsymbol{x}^n)$. Note that the algorithm's complexity scales gracefully, as it grows linearly with the length of $\boldsymbol{x}^n$. Moreover, the algorithm does not need any information about potential attack, only requiring knowledge of the signals statistics as given by $F_w^\Lambda$.%, which is needed to compute $\lambda(\cdot,\cdot)$. %The problem of learning the signal statistics is interesting, but is left for future work.
\begin{algorithm}
\caption{Computation of $\Lambda_{\boldsymbol{X}^n}(\boldsymbol{x}^n)$}\label{alg:social}
\begin{algorithmic}[1]
\Function{Loglikelihood}{$\boldsymbol{x}^{n},\tau$}
%\State $\text{Choose }\tau\in\mathbb{R}.$
\State $L_1 = \lambda(x_1,\tau). $
\For{$k=2, \dots, n$}
\State $L_k = L_{k-1} + \lambda(x_{k+1},\tau -L_{k-1} ). $
\EndFor
\State \Return $L_n$
\EndFunction
\end{algorithmic}
\end{algorithm}

%\State $\Lambda_{\boldsymbol{X}^2}(\boldsymbol{x}^2) = \Lambda_{X_1}(x_1) + \lambda(x_2,\tau - \Lambda_{X_1}(x_1)). $
%\State $\Lambda_{\boldsymbol{X}^3}(\boldsymbol{x}^3) = \dots \nonumber $
%\State $\dots \nonumber $

%\begin{algorithm}
%\caption{Computation of $\Lambda_{\boldsymbol{X}^n}(\boldsymbol{x}^n)$}\label{alg:social}
%\begin{algorithmic}[1]
%\Function{Loglikelihood}{$\boldsymbol{x}^{n},\tau$}
%\State $\text{Choose }\tau\in\mathbb{R}.$
%\State $\Lambda_{X_1}(x_1) = \lambda(x_1,\tau). $
%\For{$k=1, \dots, n-1$}
%\State $\Lambda_{\boldsymbol{X}^{k+1}}(\boldsymbol{x}^{k+1}) = \Lambda_{\boldsymbol{X}^{k}}(\boldsymbol{x}^{k})+ \lambda(x_{k+1},\tau -\Lambda_{\boldsymbol{X}^{k}}(\boldsymbol{x}^{k}) ). $
%\EndFor
%\State \Return $\Lambda_{\boldsymbol{X}^{n}}(\boldsymbol{x}^{n})$
%\EndFunction
%\end{algorithmic}
%\end{algorithm}

\subsection{Information cascades as strength or weakness}
\label{secC}

%It is intuitive that t
The term ``social learning'' refers to the fact that the accuracy of $X_n$ as a predictor of $W$ grows with $n$, and hence $n_\text{c}$ is usually chosen as one of the last nodes in the decision sequence. However, as the number of shared signals grows the increasing ``social pressure'' can make the nodes to ignore their individual measurements and blindly follow the dominant choice~\cite{bikhchandani1992theory}. This phenomenon, known as \textit{information cascade}, introduces severe limitations in the achievable asymptotic performance of social learning~\cite{acemoglu2011bayesian}.

A positive effect of information cascades, which has been overlooked before, is to make a large number of agents/nodes to hold equally qualified estimator(s), generating many locations where the network operator can collect and aggregate the data. This property avoids the existence of a single point of failure to robustly face topology-aware attacks. An attempt to blindly guess $n_\text{c}$ in order to tamper the $n_\text{c}$-node would be inefficient due to the large number of potential candidates.

However, an attacker can also leverage the information cascade phenomenon. A rational attacking strategy is to tamper the first $N^*$ nodes of the decision sequence, setting their signals in order to push the networked decisions towards a misleading cascade\footnote{Intuitively, it is more likely for a node to follow a misleading cascade if all the previous $N^*$ nodes have been tampered and act homogeneously, than for a node of higher index if the previous decisions are non-homogeneous. %A rigorous proof of this is part of our ongoing work.
}. If $N^*$ is large enough an information cascade can be triggered almost surely, making the learning process to fail. However, if $N^*$ is not large enough then the network may undo the initial pool of wrong opinions and end up triggering a correct cascade anyway. This capability of ``resilience'' %depends on the signals distribution, and 
is explored in the next section.

\section{Proof of concept}
\label{sec:4}

%\subsection{Scenario description}

To illustrate the application of social learning against topology-aware data falsification attacks, we consider a network of randomly distributed sensors over a sensitive area following a Poisson Point process (PPP). The ratio of the area that is within the range of each sensor is denoted by $r$. If attacks occur uniformly over the surveilled area, then $r$ is also the probability of an attack taking place under the coverage area of a particular sensor is. It is further assumed that each node is equipped with a binary sensor (i.e. $S_n\in\{0,1\}$), whose probability of generating a wrong measurement due to electronic and other imperfections is denoted by $q$.

For finding the posterior distributions of $S_n$, first note that $\Pw{0}{ S_n =1} = q$,
%
%\begin{equation}
%\Pw{0}{ S_n =1} = q, \nonumber
%\end{equation}
%
as a sensor false-alarm can only be due to noise. The probability of detecting an event is given by
\begin{align}
\mathbb{P} \{ S_n = 1 | W=1& \} = \P{ \text{\footnotesize{attack in range, good measurement}}|W=1} \nonumber\\
& + \P{ \text{\footnotesize{attack out of range, bad measurement}}|W=1} \nonumber\\
%&= r(1-q) + (1-r)q \nonumber\\
&= r + q - 2rq  \nonumber
\enspace.
\end{align}
Therefore, the sensor miss-detection rate is $\Pw{1}{S_n = 0}  = 1 - r -q + 2rq$. The signal log-likehood is hence given by
\begin{equation}\label{eq:signaaaa}
\Lambda_S(S_n) = 
S_n \log \frac{ r+q-2rq }{ q }  + ( 1 - S_n ) \log \frac{ 1 - r - q + 2rp }{ 1-q } 
.\nonumber
\end{equation}
Note that $\Lambda_S(1) > \Lambda_S(0)$, which is consequence of $r+q-2rq > q$ and $q<1/2$. Correspondingly, the c.d.f. of $\Lambda_S$ is%for given $W=w$ is
\begin{equation}
F_w^\Lambda(l) = 
\begin{cases}
0 \qquad &\text{if $l< \Lambda(0)$,} \\
\P{S_n=0|W=w} &\text{if $\Lambda(0) \leq l < \Lambda(1)$,} \\
1 \     &\text{if $\Lambda(1) \geq l$.}
\end{cases} \nonumber
\end{equation}
%
%Hence, each node tries %decision is The inference problem is hence 
%to distinguish between two Bernoulli variables, one with parameter $q$ and the other with parameter $r+q-2rq$. Note that, as hypothesis $W=1$ is more likely to generate a $1$ that hypothesis $W=0$, then the only non-trivial strategy based on a single measurement is to choose $X_n=S_n$. However, if $r=5\%$ and $q=10^{-3}$ this strategy give a miss-detection rate of $0.949$, indicating that without collaboration each node is very unreliable.

%\subsection{Simulation results}

We studied a network composed by $N=200$ sensor nodes, generating $\boldsymbol{X}^n$ sequentially following \eqref{eqL21344315} and using Algorithm~\ref{alg:social} to compute $\Lambda_{\boldsymbol{X}^n}(\boldsymbol{X}^n)$. Following Section~\ref{secC}, it is assumed that a topology-aware attacker tampered the first $N^*$ nodes of the decision sequence and uses them to increase the miss-detection rate by setting $X_n=0$ for $n=1,\dots, N^*$. %, increasing the miss-detection rate. 
Finally, in order to favour the reduction of miss-detections over false alarms, $\tau=0$ is chosen as is the lowest value that still allows a non-trivial inference process.%\footnote{Simulations showed that if $\tau<0$ then $X_n=1$ for all $n\in\mathbb{N}$ independently of the value of $W$, triggering a premature information cascade.}. 
For each set of parameter values, $10^4$ simulation runs are performed. 

Simulations demonstrate that the proposed scheme enables strong network resilience in this scenario, allowing the sensor network to maintain a low miss-detection rate even in the presence of an important number of Byzantine nodes (see Figure~\ref{fig:nodebynode}). In contrast, f a traditional distributed detection scheme is used, a topology-aware attacker can cause a miss-detection rate of $100\%$ by just compromising the few nodes that perform data aggregation, i.e. the FC(s). Figure~\ref{fig:nodebynode} shows that nodes aggregating data by social learning can achieve an average asymptotic miss-detection rate of less than $5\%$ even when $30\%$ of the most critical nodes are under the control of the attacker, having some resemblance with the well-known 1/3 threshold of the Byzantine generals problem~\cite{lamport1982byzantine}. Moreover, Figure~\ref{fig:nodebynode} also suggest that our scheme can still provide network resilience within the $10\%$ most unfavorable cases. %These results confirms that our data aggregation scheme effectively avoids having single points of failure. 
% by making all the nodes to agregate data, the network can overcome the influence of Bizantine nodes
%and hence even when some nodes have been compromised the rest of the network can
% and generate a correct inference.
%
%\vspace{-3cm}
\fig{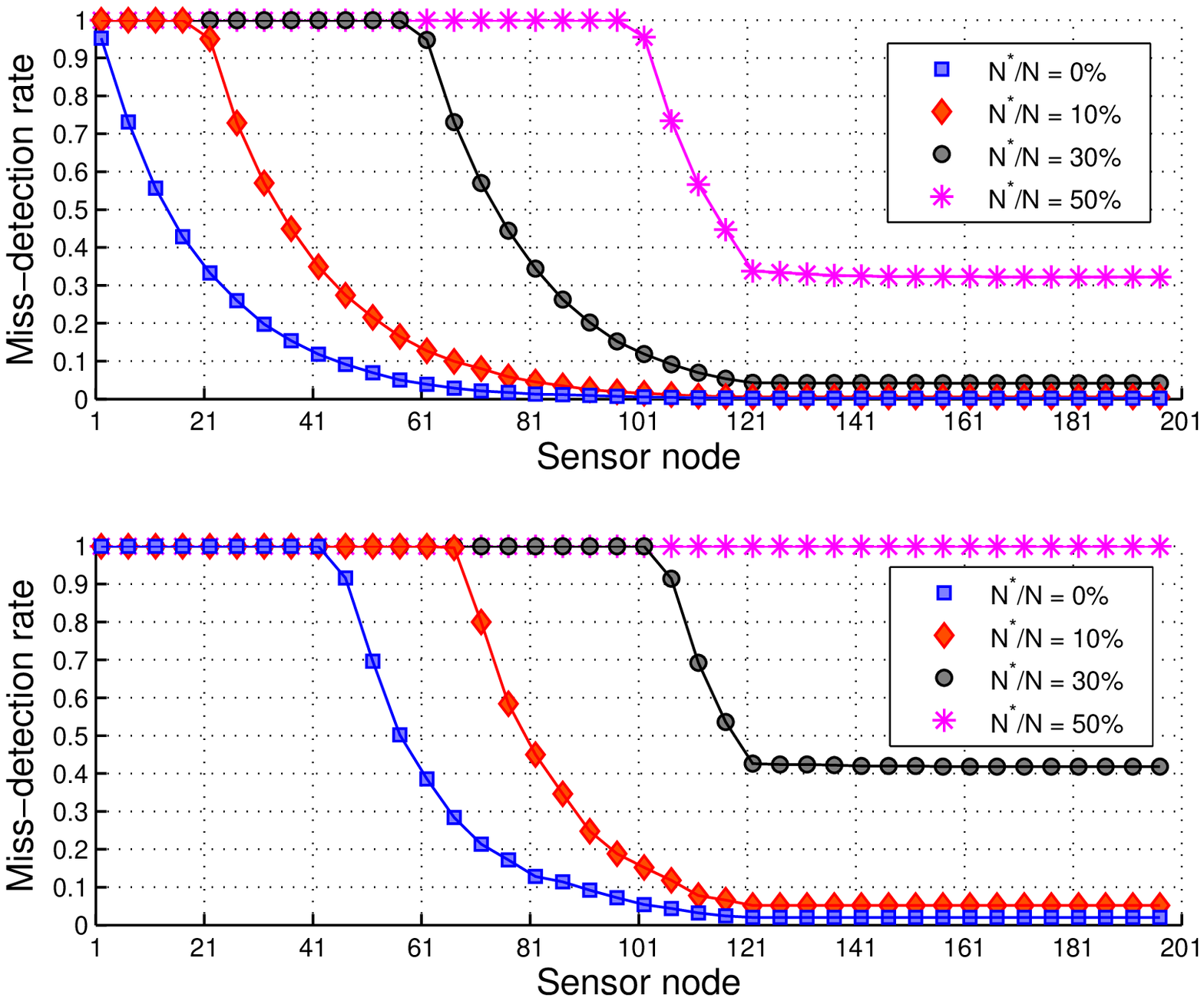}{fig:nodebynode}{\textit{Above:} Performance of a surveillance network based on social learning, with binary signals of range $r=5\%$ and error rate $q=10^{-4}$, when $N^*$ out of $N$ nodes are compromised by an attacker. \textit{Bellow:} Performance considering the $10\%$ most unfavorable cases.}
%In this scenario social learning achieves a low asymptotic miss-detection rate even when $30\%$ of the nodes are Bizantine.

Interestingly, the data aggregation is performed node by node independently of the network size. Hence, in a very large network the first 200 nodes would exhibit the same performance as the one shown in Figure~\ref{fig:nodebynode}. Adding more nodes %to the network 
may not introduce significant improvements to the asymptotic performance, as the asymptotic estimator is 
%practically attained already by the 150-th node, being
copied by later nodes following an information cascade. Nevertheless, in a large network information cascades provide the fundamental benefit of creating a large number of nodes from where the network operator can access aggregated data.

The network resilience provided by our scheme is influenced by the sensor statistics, which are determined by $q$ and $r$ (see Figure~\ref{fig:asy}). Intuitively, the achievable miss-detection rate under a low number of Byzantine nodes is reduced by a smaller $q$ or larger $r$. Furthermore, our numerical results suggest that the number of Byzantine nodes affects the miss-detection rate exponentially with a rate of growth inversely proportional to $r$, as nodes with smaller $r$ trust each other’s decisions less and hence are less affected by ``social pressure''. Consequently, it is desirable to deploy sensors with smaller probability of malfunction ($q$) than larger coverage ($r$), as a larger coverage makes the network more vulnerable to Byzantine nodes and subsequent misleading information cascades.
\fig{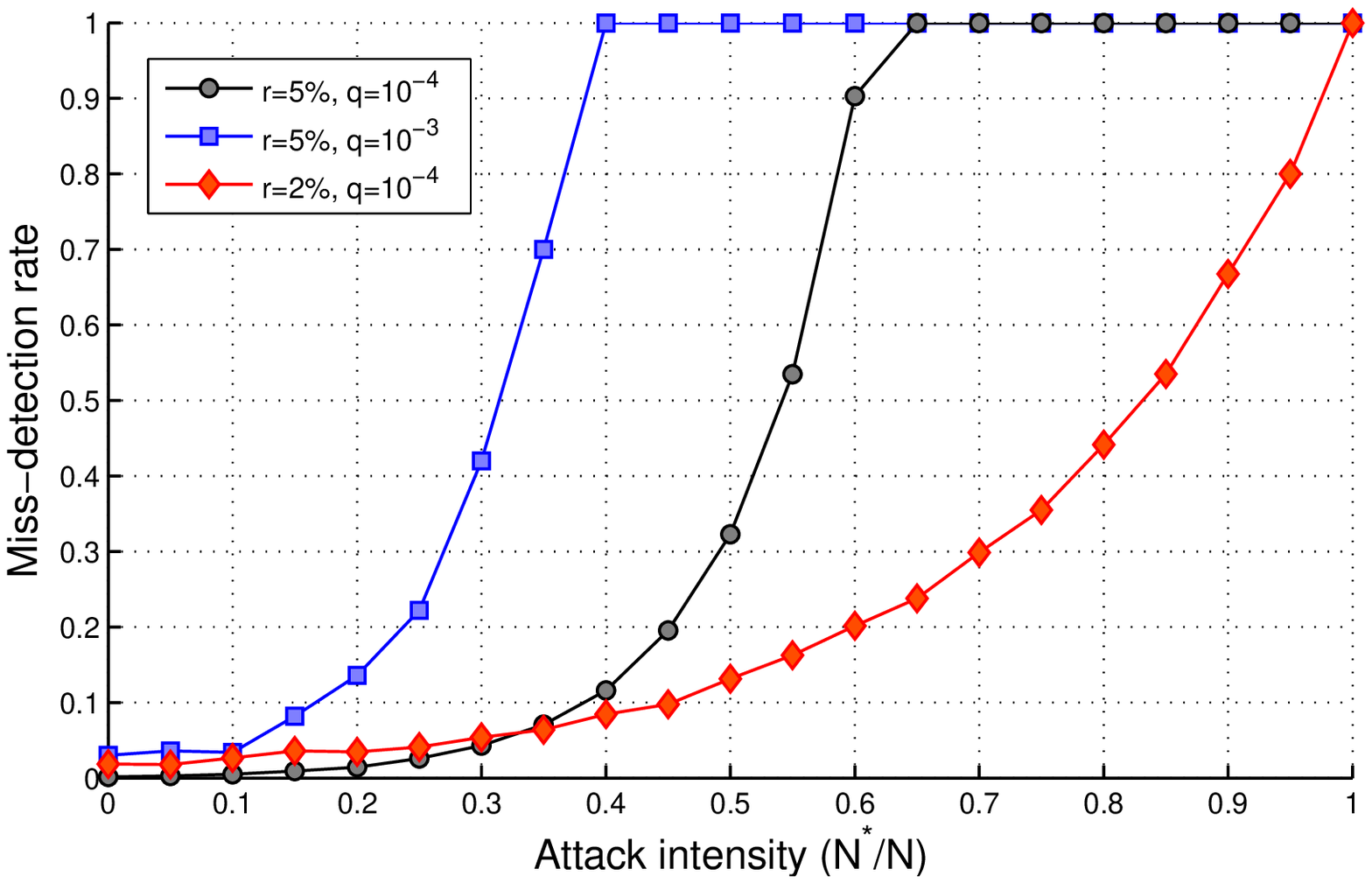}{fig:asy}{Asymptotic average performance of a surveillance system. A smaller sensor error rate ($q$) or large sensing range ($r$) improves the performance under a low $N^*$, but the latter also makes the performance degradation less graceful when $N^*$ grows.}

Our scheme does not require knowledge about attack statistics, being well-suited for practical scenarios as operation in large scale or mobile scenarios suggest dynamically changing topology. Moreover, simulations show that if the adversary tamper not the initial nodes but a different set of the same cardinality, then the attack has less impact over the system performance. This suggests that our scheme can provide further resilience against attackers who are not topology-aware. %The only requirement for setting up 

% ---------------------------------------------------------------- %
% --- APPENDICES ------------------------------------------------- %
% ---------------------------------------------------------------- %
\appendices
%\input{\textpath/appendix7}
%\input{\textpath/appendix1}
%\input{\textpath/appendix2}
%\input{\textpath/appendix3}
%\input{\textpath/appendix4}
%\input{\textpath/appendix5}
%\input{\textpath/appendix9}
%\input{\textpath/appendix6}
%\input{\textpath/appendix8}
%\section*{Acknowledgment}

%...

%This work was funded by grant FONDECYT 1110370 and by ``Programa de Cooperaci\'on Cient\'ifica Internacional CONICYT/DAAD'', which enabled an enriching exchange with Dr. Volker Jungnickel and Konstantinos Manolakis from the Fraunhofer Heinrich-Hertz-Institut in Berlin, Germany.

% ---------------------------------------------------------------- %
% --- REFERENCES ------------------------------------------------- %
% ---------------------------------------------------------------- %
\bibliographystyle{IEEEtran}
\balance
%\bibliography{sections/ref}
\bibliography{refs}

% ---------------------------------------------------------------- %
% --- FIGURES AT END --------------------------------------------- %
% ---------------------------------------------------------------- %
%\ifCLASSOPTIONpeerreview  % Peer-review mode
%  \clearpage

%\fig{SignalP_t.eps}{fig:SigIntP_t}{Mean values of the signal power \eqref{SignalP_t} and interference power \eqref{InterferenceP_t} over time (solid lines) for a downlink cellular system with $N_b = 7$ coordinated base stations and $N_u$ users, with perfect initial channel estimation and channel variation over time. Markers indicate simulation results.}

%\clearpage
%\fig{SIR.eps}{fig:SIR}{Relationship between number of users and SIR over time \eqref{SIR_t} of a downlink cellular system with coordinated base stations ($N_b=7$) and perfect channel knowledge.}

%\fi

% ---------------------------------------------------------------- %
% --- BIOGRAPHIES ------------------------------------------------ %
% ---------------------------------------------------------------- %
\ifCLASSOPTIONjournal % Conference papers do not normally have biographies
\fi

\end{document}